\documentclass[runningheads]{llncs}

\usepackage{amssymb}
\usepackage{graphicx}
\usepackage{epstopdf}
\usepackage{amsmath}
\usepackage{algorithm} 
\usepackage{algpseudocode}
\usepackage{array}
\usepackage{multirow}

\newcommand*{\Scale}[2][4]{\scalebox{#1}{$#2$}}%
\newcommand{\pair}[2]{\Scale[0.82]{\begin{pmatrix} #1 \\ #2 \end{pmatrix}}}

\makeatletter
\newcommand*{\rom}[1]{\expandafter\@slowromancap\romannumeral #1@}
\makeatother

\begin{document}

\mainmatter  % start of an individual contribution

\title{Fast Multiple Order-Preserving Matching Algorithms}

\titlerunning{Fast Multiple Order-Preserving Matching Algorithms}

\author{Myoungji Han\inst{1} \and Munseong Kang\inst{2} \and Sukhyeun Cho\inst{2} \and Geonmo Gu\inst{1}, Jeong Seop Sim\inst{2} \and Kunsoo Park\inst{1}}
\authorrunning{M. Han et al.}

\institute{Department of Computer Science and Engineering, Seoul National University \\
\email{\{mjhan,gmgu,kpark\}@theory.snu.ac.kr} \and
Department of Computer Science and Information Engineering, Inha University \\
\email{\{kmsung1125,csukhyeun\}@inha.edu, jssim@inha.ac.kr} \\
}


\toctitle{Lecture Notes in Computer Science}
\tocauthor{Authors' Instructions}
\maketitle

\begin{abstract}
Given a text $T$ and a pattern $P$, the order-preserving matching problem is to find all substrings in $T$ which have the same relative orders as $P$. Order-preserving matching has been an active research area since it was introduced by Kubica et al. \cite{kubica2013linear} and Kim et al. \cite{kim2014order}. In this paper we present two algorithms for the multiple order-preserving matching problem, one of which runs in sublinear time on average and the other in linear time on average. Both algorithms run much faster than the previous algorithms.
%\keywords{order-preserving matching, multiple pattern matching}
\end{abstract}

\section{Introduction}\label{Introduction}
\begin{figure}
\begin{center}\
  \centerline{\includegraphics[scale=0.7]{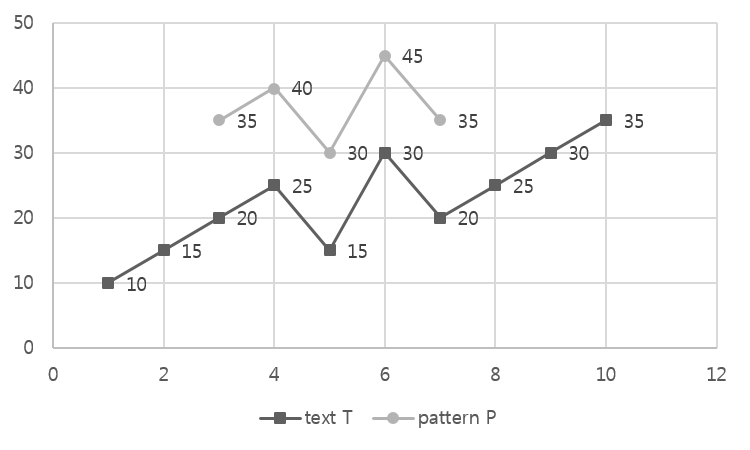}}
  \caption{An example of the order-preserving matching. $P$ has the same relative orders as the substring $T'=(20,25,15,30,20)$ of $T$. }
  \label{fig:opmex}
  \end{center}
\end{figure}

Given a text $T$ and a pattern $P$, the order-preserving matching problem is to find all substrings in $T$ which have the same relative orders as $P$. For example, given $T=(10,15,20,25,15,30,20,25,30,35)$ and $P=(35,40,30,45,35)$, $P$ has the same relative orders as the substring $T'=(20,25,15,30,20)$ of $T$. In $T'$ (resp. $P$), the first character $20$ (resp. $35$) is the second smallest number, the second character $25$ (resp. $40$) is the third smallest number, the third character $15$ (resp. $30$) is the smallest number, and so on. See Fig.~\ref{fig:opmex}. This problem is naturally generalized to the problem of finding multiple patterns. The order-preserving matching for a single pattern will be called the \emph{single order-preserving matching}, and one for multiple patterns the \emph{multiple order-preserving matching}. In this paper we are concerned with the multiple order-preserving matching problem.

Order-preserving matching was introduced by Kubica et al. \cite{kubica2013linear} and Kim et al. \cite{kim2014order}, where Kubica et al. \cite{kubica2013linear} defined order relations by order isomorphism of two strings, while Kim et al. \cite{kim2014order} defined them explicitly by the sequence of rank values, which they called the \emph{natural representation}. They both proposed $O(n+m\log{m})$ time solutions for the single order-preserving matching based on the Knuth-Morris-Pratt algorithm, where $n$ is the length of the text and $m$ is the length of the pattern. Kim et al. \cite{kim2014order} also proposed an $O(n\log{M})$ time algorithm for the multiple order-preserving matching based on the Aho-Corasick algorithm, where $M$ is the sum of lengths of all the patterns. Henceforth, there has been considerable research on the single and multiple order-preserving matching problems. For the single order-preserving matching, Cho et al. \cite{cho2015fast} proposed a method to apply the Boyer-Moore bad character rule to order-preserving matching by using the notion of $q$-grams. Chhabra and Tarhio \cite{chhabra2014order} presented a more practical solution based on filtering. They first encoded input sequences into binary sequences and then applied standard string matching algorithms as a filtering method. Faro and K\"ulekci~\cite{faro2015efficient} improved Chhabra and Tarhio's solution by using new encoding techniques which reduced the false positive rate of the filtering step. For the multiple order-preserving matching, Belazzougui et al. \cite{belazzougui2013single} theoretically improved the solution of Kim et al. \cite{kim2014order} by replacing the underlying data structure by the van-Emde-Boas tree. They achieved randomized $O(n\cdot \text{min}(\log{\log{n}},\sqrt{\frac{\log{r}}{\log{\log{r}}}},k))$ time for the search, where $r$ is the length of the longest pattern and $k$ is the number of patterns. 
%But it is well known that the Van-Emde-Boas tree is not practical due to its constant overhead. 

Order-preserving matching has been an active research area and many related problems have been studied such as order-preserving suffix trees \cite{crochemore2013order} and order-preserving matching with $k$ mismatches \cite{gawrychowski2014order}. Kim et al. \cite{kim2014representations} extended the representations of order relations from binary relations to ternary relations. With their representations, one can modify earlier order-preserving matching algorithms to accommodate strings with duplicate characters, i.e., a number can appear more than once in a string.

In this paper, we present two new algorithms for the multiple order-preserving matching problem which are more efficient on average than the previously proposed algorithms. The algorithms are based on modifications of some conventional pattern matching algorithms such as Wu-Manber \cite{wu1994fast} and Karp-Rabin \cite{karp1987efficient}. The first algorithm, called Algorithm \rom{1}, uses the ideas of the Wu-Manber algorithm, and the second algorithm, called Algorithm \rom{2}, uses the ideas of the Karp-Rabin algorithm and the encoding techniques of Chhabra and Tarhio \cite{chhabra2014order} and Faro and K\"ulekci \cite{faro2015efficient} for fingerprinting. Algorithm \rom{1} runs in $O(\frac{n}{m}\log{M})$ time on average, where $n$ is the length of the text, $m$ is the length of the shortest pattern, and $M$ is the sum of lengths of all the patterns. Algorithm \rom{2} runs in $O(n)$ time on average, assuming that $M$ is polynomial with respect to $m$. 
In order to verify practical behaviors of our algorithms, we conducted experiments where the two algorithms were compared with the algorithms of Kim et al. \cite{kim2014order} and Belazzougui et al. \cite{belazzougui2013single}. Experiments show that our algorithms run much faster in practice.

%This paper is organized as follows. In Section~\ref{section:pf} we define the problem to be solved and introduce some definitions related to order-preserving matching. In Section \ref{section:alg1} and \ref{section:alg2} we present our new algorithms and analyze their time complexities. In Section~\ref{section:exp} we give experimental results and discuss them. Finally, Section~\ref{section:con} concludes the paper.

\section{Problem Formulation}\label{section:pf}

Let $\Sigma$ denote a set of numbers such that a comparison of two numbers can be done in constant time, and let $\Sigma^*$ denote the set of strings over the alphabet $\Sigma$. For a string $x \in \Sigma^*$, let $|x|$ denote the length of $x$. A string $x$ is described by a sequence of characters $(x[1], x[2], ..., x[|x|])$.
Let a substring $x[i..j]$ be $(x[i], x[i+1], ..., x[j])$ and a prefix $x_i$ be $x[1..i]$.
For a character $c \in \Sigma$, let $rank_x(c) = 1 + |\{ i : x[i] < c \; \text{for} \; 1 \leq i \leq |x| \}|$.

We use the \emph{natural representation} defined by Kim et al.~\cite{kim2014order} to compare order relations of two strings. The natural representation is equivalent to \emph{order-isomorphism} defined by Kubica et al.~\cite{kubica2013linear}, because the natural representation of two strings are identical if and only if they are order-isomorphic.
\begin{definition}[Natural representation \cite{kim2014order}] For a string $x$ of length $n$, the natural representation is defined as $Nat(x) = ( rank_x(x[1]), rank_x(x[2]) , ..., rank_x(x[n]) )$.
\end{definition}
For example, for $x=(30, 40, 30, 45, 35)$, the natural representation is
$Nat(x)=(1, 4, 1, 5, 3)$. We will simply say that $x$ matches $y$ if $|x|=|y|$ and $Nat(x)=Nat(y)$.

Order-preserving matching can be defined in terms of the natural representation.

\begin{definition}[Single Order-Preserving Matching \cite{kim2014order}]\label{def:sopm}
Given a text $T[1..n] \in \Sigma^*$ and a pattern $P[1..m] \in \Sigma^*$, $P$ matches $T$ at position $i$ if
$Nat(P)=Nat(T[i-m+1..i])$. The order-preserving matching is the problem of finding all the positions of $T$ matched with $P$.
\end{definition}
Definition~\ref{def:sopm} is naturally generalized to the multiple order-preserving matching, formally defined in Definition~\ref{def:mopm}.

\begin{definition}[Multiple Order-Preserving Matching \cite{kim2014order}]\label{def:mopm}
Given a text $T[1..n] \in \Sigma^*$ and a set of patterns $\mathcal{P}=\{ P_1, P_2, ..., P_k \}$ where $P_i \in \Sigma^*$ for $1 \leq i \leq k$, the multiple order-preserving matching is the problem of finding all the positions of $T$ matched with any pattern in $\mathcal{P}$.
\end{definition}

There are two other representations in addition to the natural representation for comparing order relations of two strings: \emph{prefix representation} and \emph{nearest neighbor representation}. The \emph{prefix representation} can be defined as a sequence of rank values of characters in prefixes.

\begin{definition}[Prefix Representation~\cite{kim2014order}] For a string $x$, the prefix representation is defined as $Pre(x) = ( rank_{x_1}(x[1]), rank_{x_2}(x[2]) , ..., rank_{x_{|x|}}(x[|x|]) )$.
\label{appendix:definition:prefix_rep}
\end{definition}
For example, for $x=(30, 40, 30, 45, 35)$, the prefix representation is
$Pre(x)=(1, 2, 1, 4, 3)$. We can compute $Pre(x)$ in time $O(|x|\log{|x|})$ for general alphabet using the order-statistic tree~\cite{kim2014order}. The time complexity can be reduced to $O(|x|)$ if the characters can be sorted in $O(|x|)$ time.

\begin{lemma}~\cite{cho2015fast} 
For two strings $x$ and $y$ where $|x|=|y|$, if $x$ matches $y$, then $Pre(x)=Pre(y)$.
\label{lemma:if}
\end{lemma}

The prefix representation has an ambiguity between different strings if they include duplicate characters. For example, when $x=(10,30,20)$, and $y=(10,20,20)$, the prefix representations of both $x$ and $y$ are $(1,2,2)$, whereas their natural representations are different. Kim et al. defined a new representation called the \emph{extended prefix representation}~\cite{kim2014representations} for strings with duplicate characters. We omit the details here.

For the \emph{nearest neighbor representation}, we define $LMax_x[i]$ and $LMin_x[i]$ as follows.
\[ LMax_x[i] = \left\{
  \begin{array}{l l}
    j & \quad \text{if $x[j] = \max \{ x[k]: x[k] \leq x[i] \text{ for } 1 \leq k \leq i-1 \}$ }\\
    -\infty & \quad \text{if no such $j$ ,}
  \end{array} \right.\]
\[ LMin_x[i] = \left\{
  \begin{array}{l l}
    j & \quad \text{if $x[j] = \min \{ x[k]: x[k] \geq x[i] \text{ for } 1 \leq k \leq i-1 \}$ }\\
    \infty & \quad \text{if no such $j$ .}
  \end{array} \right.\]
If there are multiple $j$'s for $LMax_x[i]$ or $LMin_x[i]$, we choose the rightmost one.

\begin{definition}[Nearest Neighbor Representation~\cite{kim2014representations,kim2014order}] For a string $x$, the nearest neighbor representation is defined as $\text{NN}(x) =$
$\pair{LMax_{x}[1]}{LMin_{x}[1]}$
$\pair{LMax_{x}[2]}{LMin_{x}[2]}$
$\cdots$
$\pair{LMax_{x}[|x|]}{LMin_{x}[|x|]}$.
\end{definition}
For example, for $x=(30, 40, 30, 45, 30)$, the nearest neighbor representation is as follows.
\[
\begin{array}{l}
\text{NN}(x)=\left(\pair{-\infty}{\infty}, \pair{1}{\infty}, \pair{1}{1}, \pair{2}{\infty}, \pair{3}{3}\right)    \;.
\end{array}
\]
For convenience, let $x[-\infty] = -\infty$, $x[\infty] = \infty$, $Nat(x)[-\infty] = 0$ and $Nat(x)[\infty] = |x|+1$ for any string $x$. Then, $Nat(x)[LMax_x[i]] \leq Nat(x)[i] \leq Nat(x)[LMin_x[i]]$ holds for $1\le i\le |x|$.

The time complexity for computing $\text{NN}(x)$ is $O(|x|\log{|x|})$~\cite{kim2014order}. Using this representation, we can check if two strings match in time linear to the size of the input, even when the strings have duplicate characters.

\begin{lemma}~\cite{cho2015fast,kim2014representations,kim2014order,kubica2013linear} Given two strings $x$ and $y$ where $|x|=|y|$, assume $\text{NN}(x)$ is computed. Then we can determine whether $x$ matches $y$ in $O(|x|)$ time.
\label{lemma:nn matching}
\end{lemma}

\section{Algorithm \rom{1}}\label{section:alg1}
In this section, we present our first algorithm for the multiple order-preserving matching. Algorithm \rom{1} is based on the Wu-Manber algorithm, which is widely used for multiple pattern matching. Algorithm \rom{1} is divided into two steps: the preprocessing step and the searching step. 

\subsection{Preprocessing Step of Algorithm \rom{1}}
Let $m$ be the length of the shortest pattern, and $M$ be the sum of lengths of all the patterns. We consider only the first $m$ characters of each pattern. Let $\mathcal{P}'=\{P_1',P_2',\cdots ,P_k'\}$ where $P_i'=P_i[1..m]$ (this notation is provided only for clarity of exposition). In the preprocessing step, we build a SHIFT table and a HASH table based on $\mathcal{P}'$, which are analogous to those of the Wu-Manber algorithm. However, since we are looking for strings matched with patterns in terms of order-preserving matching, we have to consider the order representations of strings rather than strings themselves for comparison. Consider a block of length $b$ on the text, where $b\le m$. The SHIFT table determines the shift value based on the prefix representation of the given block. Given a block $x$, we define
\begin{eqnarray*}
\ l_x=\max\{j : Pre(P_i'[j-b+1..j])=Pre(x) \;\text{for} \; 1\le i\le k, b\le j\le m \} \;.
\end{eqnarray*}
That is, $l_x$ means the position of the rightmost block in any $P_i'\in\mathcal{P}'$ which is likely to match $x$. Here, the term "is likely to" is used because $Pre(x)=Pre(y)$ does not necessarily mean that $x$ matches $y$. For convenience, let $l_x=-\infty$ if there is no such block. Then, the SHIFT table is defined as 
\begin{eqnarray*}
\ \text{SHIFT}[f(x)]=\min(m-l_x, m-b+1)\;,
\end{eqnarray*}
where $f(x)$ is a fingerprint mapping a block $x$ to an integer used as an index to the SHIFT table. Using the factorial number system \cite{kunuth1998art}, we define $f(x)$ as
\begin{eqnarray*}
\ f(x)=\sum_{i=1}^{b} (Pre(x)[i]-1)\cdot (i-1)!\;.
\end{eqnarray*}
Note that $f(x)$ maps a block $x$ into a unique integer within the range $[0..b!-1]$ according to its prefix representation.

\begin{figure}
  \begin{center}\
  \centerline{\includegraphics[scale=0.30]{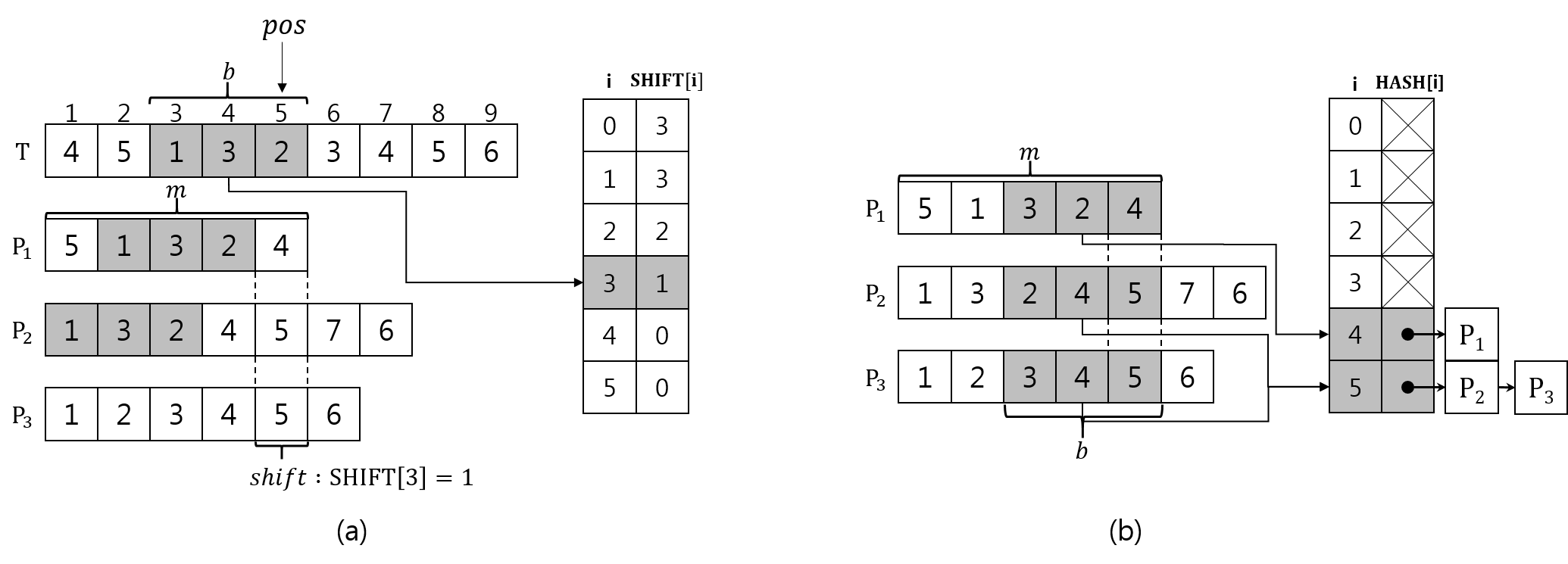}}
  \caption{SHIFT and HASH tables.}
  \label{alg1_example}
  \end{center}
\end{figure}

Fig.~\ref{alg1_example}-(a) shows the SHIFT table when there are three patterns. Assume that $b=3$. Consider the block $T[3..5]$. The rightmost block in $\mathcal{P}'$ whose prefix representation equals that of $T[3..5]$ is $P_1[2..4]$. The fingerprint $f(T[3..5])$ is $3$. Thus, $\text{SHIFT}[3]$ is $m-4=1$. Note that in the figure, we can safely shift the patterns by $1$.

The fingerprint is also used to index the HASH table. $\text{HASH}[i]$ contains a pointer to the list of the patterns whose last block in $\mathcal{P}'$ is mapped to the fingerprint $i$. Fig.~\ref{alg1_example}-(b) shows the HASH table with the same patterns.

To compute the values of the SHIFT table, we consider each pattern $P_i'$ separately. For each pattern $P_i'$, we compute the fingerprint of each block $P_i'[j-b+1..j]$ consecutively, and set the corresponding value of the SHIFT table to the minimum between its current value (initially set to $m-b+1$) and $m-j$. In order to obtain the fingerprint of a block, we have to compute its prefix representation. Once we compute the fingerprint of the first block $Pre(P_i'[1..b])$ using the order-statistic tree, the tree contains the first $b$ characters of $P_i'$. To compute the prefix representations of the subsequent blocks, we observe that we can compute $Pre(P_i'[j+1..j+b])$ by taking advantage of the order-statistic tree containing characters of the previous block $P_i'[j..j+b-1]$. Specifically, we erase $P_i'[j]$ from the tree and insert the new character $P_i'[j+b]$ into the tree. Inserting and deleting an element into the order-statistic tree is accomplished in $O(\log{b})$ time since the tree contains $O(b)$ elements. Then we traverse the tree in $O(b)$ time to retrieve the prefix representation of the new block. We repeat this until we reach the last block. When we reach the last block, we map into the HASH table and add $P_i$ into the corresponding list. The whole process is performed for all the patterns. Since there are $O(km)$ blocks, it takes $O(kmb)$ time to construct the SHIFT and HASH tables. %, which becomes $O(\frac{M\log{M}}{\log{\log{M}}})$ since $b=\frac{1.5\log{M}}{\log{\log{M}}}$.

We also precompute the nearest neighbor representations of all the patterns, namely, $\text{NN}(P_i)$ for $1\le i\le k$. They are used in the searching step for verifying whether patterns actually match the text. Using the order-statistic tree, they are computed in $O(M\log{r})$ time, where $r$ denotes the length of the longest pattern. As a result, the time complexity for the preprocessing step is $O(kmb+M\log{r})$.

\subsection{Searching Step of Algorithm \rom{1}}
In the searching step, we find all the positions of $T$ matched with any pattern in $\mathcal{P}$. Fig.~\ref{alg1} shows the pseudocode of Algorithm \rom{1}. For the search, we slide a position $pos$ along the text, reading a block of length $b$, $T[pos-b+1..pos]$, and computing the corresponding fingerprint $i$. If $\text{SHIFT}(i)>0$, then we shift the search window to $pos+\text{SHIFT}(i)$ and continue the search. Otherwise, $\text{SHIFT}(i)=0$ and there may be a match. Thus we select the list of patterns in $\text{HASH}[i]$, and compare each pattern in the list with the text via the nearest neighbor representation. We call this process the verification step. We repeat this until we reach the end of the text.

\begin{figure}

\begin{normalsize}
\textbf{Algorithm \rom{1}($P=\{P_1,P_2,\cdots,P_k\}$, $T[1..n]$)}
\end{normalsize}
\begin{algorithmic}[1]
\State $m\gets min_{1\le i\le k}(|P_i|)$
\State Preprocess $P$ and compute SHIFT, HASH, NN
\State $pos\gets m$
\While{$pos\le n$}
	\State $i\gets f(T[pos-b+1..pos])$
	\If{$\text{SHIFT}[i]=0$}
		\State Verify each pattern in $\text{HASH}[i]$ via NN
		\State $pos\gets pos+1$
	\Else
		\State $pos\gets pos+\text{SHIFT}[i]$
	\EndIf
\EndWhile
\end{algorithmic}
\caption{The pseudocode of Algorithm \rom{1}}\label{alg1}
\end{figure}

\subsection{Average Time for the Search of Algorithm \rom{1}} \label{anal1}
We present a simplified analysis of the average running time for the searching step. For the analysis, we assume that there are no duplicate characters in any $b$-length block in strings, i.e., any consecutive $b$ characters in the text and patterns are distinct. Although this assumption restricts the generality of our problem, it is insignificant because: (1) a fairly large alphabet makes the case against the assumption very unlikely to happen; (2) even if it happens, the algorithm still works correctly without a significant impact on the performance in practice. We leave it as an open problem whether the average $O(\frac{n}{m}\log{M})$ time can be derived when the strings are totally random, which is more complicated. Now, we assume that each distinct block appears randomly at a given position (i.e., with the same probability). Let us denote $\sigma=|\Sigma|$, then there are $_\sigma \!P_b$ different possible blocks and the probability of a block to appear is $1/{_\sigma \!P_b}$.

\begin{lemma}
\label{lemma1}
For two random blocks $x$ and $y$, where $x,y\in {\Sigma}^b$ and each has no duplicate characters, the probability that $Pre(x)=Pre(y)$ is $\frac{1}{b!}$.
\end{lemma}

% We assume that both the text and patterns are random strings with uniform distribution, i.e., each character occurs at a given position independently with the same probability. Note that there are $b!$ entries in the SHIFT table. We reasonably assume that the probability that a random block $x$ maps into any entry of SHIFT is $\frac{1}{b!}$.

\noindent Recall that Algorithm \rom{1} determines a shift value according to the prefix representation of a current block on the text.

\begin{lemma}
\label{lemma:shift}
The probability that a random block $x$ leads to a shift value of $j$, $0\le j\le m-b$, is at most $\frac{k}{b!}$. 
\end{lemma}

\begin{lemma}
\label{lemma5}
The expected value of a shift during the search is at least $(m-b+1)\lbrace 1-\frac{k(m-b+2)}{2b!}\rbrace$.
\end{lemma}

%Recall that all the entries of SHIFT was initialized to $m-b+1$. This fact leads to the following corollaries.
%\begin{corollary}
%The probability that a random block $x$ of size $b$ leads to a shift value $m-b+1$ is at least $1-(m-b+1) \frac{k}{b!}$.
%\end{corollary}

%\begin{corollary}
%The expected amount of a shift during the search is at least $(m-b+1)(1-(m-b+1) \frac{k}{b!})$.
%\end{corollary}

We set $b=1.5\log{M}/\log{\log{M}}$. Then, by Stirling's approximation~\cite{abramowitz1972handbook}, we can easily prove that $b!=2^{b\log{b}+b\log{e}+O(\log{b})}=\Omega(M)$, and thus the expected value of a shift is at least $\Theta(m)$. Consequently, the average number of iterations of the \textbf{while} loop during the search is bounded by $O(\frac{n}{m})$. At each iteration, we compute a fingerprint and the computation takes $O(b\log{b})=O(\log{M})$ time. Lemma~\ref{avgcost} shows that the verification step at each iteration is accomplished in constant time on average.

\begin{lemma}
\label{avgcost}
The average cost of the verification step at each iteration is $O(1)$.
\end{lemma}
\noindent Hence, the average time complexity of the searching step is roughly $O(\frac{n}{m} \log{M})$.

\section{Algorithm \rom{2}}\label{section:alg2}
In this section, we present a simple algorithm that achieves average linear time for search. Algorithm \rom{2} exploits the ideas of the Karp-Rabin algorithm and the encoding techniques of Chhabra and Tarhio~\cite{chhabra2014order} and Faro and K\"ulekci~\cite{faro2015efficient} for fingerprinting.

\subsection{Fingerprinting in Algorithm \rom{2}}
The Karp-Rabin algorithm is a practical string matching algorithm that makes use of fingerprints to find patterns, and it is important to choose a fingerprint function such that a fingerprint should be efficiently computed and efficiently compared with other fingerprints. Furthermore, the fingerprint function should be suitable for identifying strings in terms of order-preserving matching.

Given an $m$-length pattern $P$, Chhabra and Tarhio~\cite{chhabra2014order} encode the pattern into a binary sequence $\beta(P)$ of length $m-1$, where
\[ \beta(P)[i] = \left\{
  \begin{array}{l l}
    1 & \quad \text{if $P[i]<P[i+1] $ }\\
    0 & \quad \text{otherwise.}
  \end{array} \right.\]
We consider the fingerprint $\beta(P)$ as an $(m-1)$-bit binary number. We can compute $\beta(P)$ in time $O(m)$.

As $m$ increases, the fingerprint $\beta(P)$ may be too large to work with; we need at least $(m-1)$ bits to represent a fingerprint. To address this issue, we compute the fingerprints as residues modulo a prime number $p$. According to \cite{karp1987efficient}, we choose the prime $p$ pseudorandomly in the range $[1..mn^2]$. With this choice, it is proved that the probability of a single false positive due to the modulo operation while searching is bounded by $2.53/n$, which is negligibly small for sufficiently large $n$~\cite{karp1987efficient}.

Faro and K\"ulekci~\cite{faro2015efficient} proposed more advanced encoding techniques such as $q$-NR and $q$-NO. Instead of comparing between only a pair of neighboring characters, they compared between a set of $q$ characters for computing the relative position of a character. We can compute fingerprints using those techniques similarly to above. We implemented Algorithm \rom{2} using three encoding techniques, including Chhabra and Tarhio's binary encoding \cite{chhabra2014order}, $q$-NR, and $q$-NO~\cite{faro2015efficient}, for fingerprinting. In the following sections, we will describe the algorithm assuming the binary encoding.

\subsection{Preprocessing Step of Algorithm \rom{2}}
\begin{figure}
  \begin{center}\
  \centerline{\includegraphics[scale=0.35]{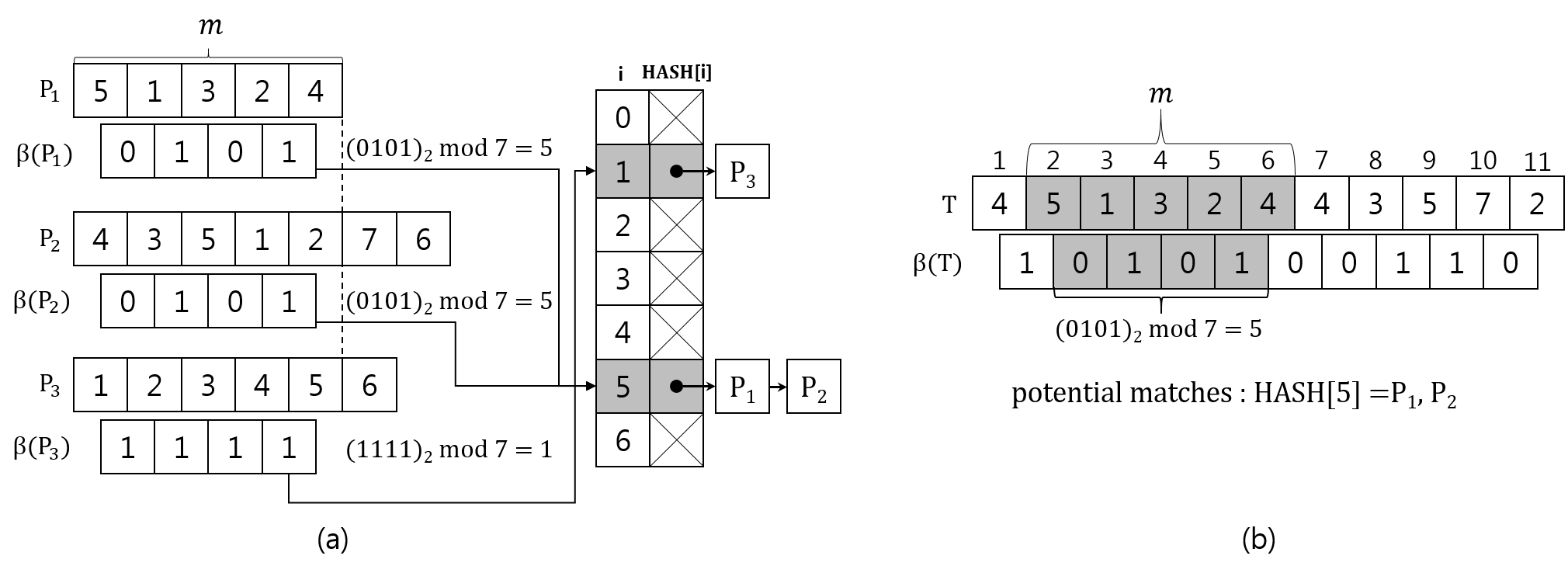}}
  \caption{(a) The HASH table. (b) An example of the search. For the window $T[2..6]$, the corresponding fingerprint is $5$. We check $\text{HASH}[5]$, which has $P_1,P_2$ as elements, and thus verify them via NN.}
  \label{alg2_example}
  \end{center}
\end{figure}

Again, let $\mathcal{P}'=\{P_1',P_2',\cdots ,P_k'\}$ be the set of $m$-length prefixes of the patterns. In the preprocessing step, we first compute $\beta(P_i')$ for $1\le i\le k$ and build a HASH table. $\text{HASH}[i]$ contains a pointer to the list of the patterns whose fingerprints equal $i$. We also compute $\text{NN}(P_i)$ for $1\le i\le k$. In total, the preprocessing step takes $O(M\log{r})$ time. Fig.~\ref{alg2_example}-(a) shows the HASH table when there are three patterns. We use a prime $p=7$ in the example. 

\subsection{Searching Step of Algorithm \rom{2}}
In the searching step, we scan the text $T$ while iteratively computing fingerprints of the successive windows of size $m$. Fig.~\ref{alg2} shows the pseudocode of Algorithm \rom{2}. We slide a search window $T[i..i+m-1]$ along the text, computing the corresponding fingerprint $\beta$. If the list pointed by $\text{HASH}[\beta]$ is not empty, we compare each pattern in the list with the text via its nearest neighbor representation. We call this process the verification step. We repeat this until we reach the end of the text. Fig.~\ref{alg2_example}(b) shows an example of the searching step.

\begin{figure}
\begin{normalsize}
\textbf{Algorithm \rom{2}($P=\{P_1,P_2,\cdots,P_k\}$, $T[1..n]$)}
\end{normalsize}
\begin{algorithmic}[1]
\State $m\gets min_{1\le i\le k}(|P_i|)$
\State Preprocess $P$ and compute HASH, NN
\State $pos\gets m$
\For{$i=1 \;\text{for}\;n-m+1$}
	\State $\beta=\beta(T[i..i+m-1]) \; \text{mod} \; p$
	\State Verify each pattern in $\text{HASH}[\beta]$ via NN
\EndFor
\end{algorithmic}
\caption{The pseudocode of Algorithm \rom{2}}\label{alg2}
\end{figure}

\subsection{Average Time for the Search of Algorithm \rom{2}}
At each iteration of the \textbf{for} loop, we compute the fingerprint $\beta$ of the search window. Let us denote $\beta_i=\beta(T[i..i+m-1]) \; \text{mod} \; p$, which is the fingerprint of the $i$-th search window. We can compute $\beta_1$ in time $O(m)$. To compute the fingerprints for the subsequent windows, we observe that we can compute $\beta_{i+1}$ from $\beta_i$ using Horner's rule~\cite{cormen2001introduction}, since
\begin{eqnarray*}
\ \beta_{i+1}=(2(\beta_i-H\cdot\beta(T)[i])+\beta(T)[i+m]) \; \text{mod} \; p\;,
\end{eqnarray*}
where $H=2^{m-2} \;(\text{mod}\;p)$ is a precomputed value. It is clear that this calculation is done in constant time.

Now, we analyze the time spent to perform the verification step. We assume that the numbers in the text and patterns are statistically independent and uniformly at random. The verification is performed when there is a match between encoded binary strings of the text and patterns. The probability that a 1 appears at a position of an encoded string is $q=({\sigma}^2/2-\sigma/2)/{\sigma}^2=(\sigma-1)/2\sigma$. So the probability of a character match~\cite{chhabra2014order} is
\begin{eqnarray*}
\ s=q^2+(1-q)^2=\frac{1}{2}+\frac{1}{2\sigma^2}\;.
\end{eqnarray*}
Since the odd positions of an encoded string are mutually independent, we can (upper) bound the probability of a match between two encoded strings by $s^{(m-1)/2}$. Note that $s\le 5/8$ for $\sigma\geq 2$.

\begin{lemma}
\label{lemma6}
When $M$ is polynomial with respect to $m$, the average cost of the verification step during the search is $O(1)$.
\end{lemma}

\noindent Hence, the average time complexity of the searching step is $O(n)$, when $M$ is polynomial with respect to $m$.

\section{Experiments}\label{section:exp}
In order to verify the practical behaviors of our algorithms, we tested them against the previous algorithms based on the Aho-Corasick algorithm: Kim et al.'s \cite{kim2014order}, and Belazzougui et al.'s~\cite{belazzougui2013single}.\footnote[1]{For the implementation of the van-Emde-Boas tree used in \cite{belazzougui2013single}, we used the source code publicly available at https://code.google.com/p/libveb/.} Kim et al.'s algorithm is denoted by \emph{KEF}, Belazzougui et al.'s by \emph{BPR}, and Algorithm \rom{1} by \emph{Alg1}. Algorithm \rom{2} is denoted by \emph{Alg2}, followed by a notation of the encoding technique adopted for fingerprinting. Specifically, \emph{Alg2\_Bin} refers to Algorithm \rom{2} with Chhabra and Tarhio's binary encoding \cite{chhabra2014order}, and \emph{Alg2\_NR2} (resp. \emph{Alg2\_NO2}) refers to Algorithm \rom{2} with the $q$-NR (resp. $q$-NO) encoding of Faro and K\"ulekci \cite{faro2015efficient} where we set $q=2$. All algorithms were implemented in C++ and run on a Debian Linux 7(64bit) with Intel Xeon X5672 processor and 32 GB RAM. During the compilation, we used the –O3 optimization option.

We tested for a random text $T$ of length $n=10^6$ searched for $k=10,50,100$ random patterns of length $m=5,10,20,50,100$, respectively. All the texts and patterns were selected randomly from an integer alphabet $\Sigma=\{1,2,\cdots ,1000\}$ (we tested for varying alphabet sizes, but we didn't observe sensible differences in the results). For each combination of $k$ and $m$, we randomly selected a text and patterns, and then ran each algorithm. We performed this 10 times and measured the average time for the searching step. Table~\ref{table1} shows the results.

Fig.~\ref{NumberOfPattern10} in Appendix shows the average search times when $k=10$. When $m$ is less than 50, \emph{Alg2\_Bin} is the best among the algorithms, achieving a speed up of about 6 times compared to \emph{KEF}, and about 14 times compared to \emph{BPR}. As $m$ increases, however, \emph{Alg1} becomes better, achieving a speed up of about 11 times compared to \emph{KEF}, and about 24 times compared to \emph{BPR}. This is due to the increase of the average shift value during the search. The reason that the average shift value increases is that since we set $b=1.5\log{M}/\log{\log{M}}$, the block size increases as $m$ increases, and thus the probability that a block appears in the patterns decreases. Fig.~\ref{NumberOfPattern50} and \ref{NumberOfPattern100} in Appendix show the average search times when $k=50,100$. They show similar trends with Fig.~\ref{NumberOfPattern10}. One thing to note is that as $k$ increases, the point of $m$ where \emph{Alg1} becomes for the first time faster than the \emph{Alg2} family increases. We attribute this to the fact that as $k$ increases, a block appears more often in the patterns, which leads to lower shift values.

\newcolumntype{C}[1]{>{\centering\let\newline\\\arraybackslash\hspace{0pt}}m{#1}}
\begin{table}\caption{Average search times with different values for $k$ and $m$.}
\begin{center}
\begin{tabular}{|C{1.0cm}|C{1.0cm}|c|C{1.5cm}|C{1.5cm}|C{1.5cm}|C{1.5cm}|C{1.5cm}|C{1.5cm}|}
\hline
\ $k$ & $m$ & & \emph{KEF} &\emph{BPR} & \emph{Alg1} & \emph{Alg2\_Bin} & \emph{Alg2\_NR2} & \emph{Alg2\_NO2}\\ \hline \hline
\multirow{5}{*}{10} & 5 & & 527.3 & 1215.1 & 274.8 & \textbf{107.6} & 164.8 & 186.5\\ \cline{2-9}
& 10 & & 544.3 & 1258.2 & 216.9 & \textbf{91.5} & 148.8 & 197.6\\ \cline{2-9}
& 20 & & 557.1 & 1254.8 & 286.5 & \textbf{88.4} & 155.4 & 194.8\\  \cline{2-9}
& 50 & & 556.2 & 1213 & \textbf{51.1} & 65.4 & 116.7 & 203\\  \cline{2-9}
& 100 & & 561.7 & 1244.8 & \textbf{56.2} & 70.4 & 123.9 & 206.9\\  \hline \hline

\multirow{5}{*}{50} & 5 & & 598 & 1227.2 & 647.8 & 234.1 & \textbf{215.3} & 310\\ \cline{2-9}
 & 10 & & 573.8 & 1238.6 & 269.6 & \textbf{100.5} & 152.3 & 194.6\\ \cline{2-9}
 & 20 & & 562.9 & 1244.2 & 308.6 & \textbf{114.8} & 187.1 & 216.4\\ \cline{2-9}
 & 50 & & 570.7 & 1239.8 & 313.5 & \textbf{113.8} & 184.5 & 226.2\\ \cline{2-9}
 & 100 & & 587.6 & 1271.8 & \textbf{55.4} & 86.1 & 150.6 & 227.6\\ \hline \hline

\multirow{5}{*}{100} & 5 & & 569 & 1291.1 & 674.4 & 395.6 & 386.3 & \textbf{307.3}\\ \cline{2-9}
 & 10 & & 629 & 1304.3 & 522.4 & \textbf{81.9} & 100.3 & 150.5\\ \cline{2-9}
 & 20 & & 589 & 1250.2 & 498.4 & \textbf{102.6} & 164 & 205.1\\ \cline{2-9}
 & 50 & & 605.3 & 1259.9 & 103.3 & \textbf{86} & 184.8 & 225.7\\ \cline{2-9}
 & 100 & & 588.9 & 1247.2 & 73.2 & \textbf{53.8} & 182 & 227.5\\ \hline

\end{tabular}
\end{center}
\label{table1}
\end{table}

\section{Conclusion}\label{section:con}
We proposed two efficient algorithms for the multiple order-preserving matching problem. Algorithm \rom{1} is based on the Wu-Manber algorithm, and Algorithm \rom{2} is based on the Karp-Rabin algorithm and exploits the encoding techniques of the previous works~\cite{chhabra2014order,faro2015efficient}. Algorithm \rom{1} performs the multiple order-preserving matching in average $O(\frac{n}{m} \log{M})$ time, and Algorithm \rom{2} performs it in average $O(n)$ time when $M$ is polynomial with respect to $m$. The experimental results show that both of the algorithms are much faster than the existing algorithms. When the lengths of the patterns are relatively short, Algorithm \rom{2} with the binary encoding performs the best due to its inherent simplicity. However, Algorithm \rom{1} becomes more efficient as the lengths of the patterns grow.

\section*{Appendix}

%\section{Proof of Lemma \ref{lemma1}}
\textbf{Proof of Lemma \ref{lemma1}.}
The probability is equivalent to the probability that when randomly choosing a block $x$, its prefix representation is the same as that of (already chosen) $y$. The sample space consists of all $_\sigma \!P_b$ possible blocks. Notice that once we choose $b$ distinct characters regardless of order, we can order them to fit in any one of the $b!$ prefix representations. It means that there are $\sigma \choose b$ blocks belonging to each prefix representation. Therefore, the probability is ${\sigma \choose b} \cdot \frac{1}{_\sigma \!P_b}=\frac{1}{b!}$.\qed

\vspace{5mm}
\noindent \textbf{Proof of Lemma \ref{lemma:shift}.}
The necessary condition for the case that $x$ leads to a shift value $j$ is that there exists a pattern $P_i'$ whose block ending at the position $m-j$ belongs to the prefix representation of $x$. Since there are $k$ patterns, the probability of the necessary condition is $\frac{k}{b!}$.\qed

\vspace{5mm}
\noindent \textbf{Proof of Lemma \ref{lemma5}.}
Since all the entries of SHIFT were initialized to $m-b+1$, the expected value of a shift is $\geq \sum_{j=0}^{m-b}j\cdot\frac{k}{b!} + (m-b+1)\lbrace 1-(m-b+1)\frac{k}{b!}\rbrace=(m-b+1)\lbrace 1-\frac{k(m-b+2)}{2b!}\rbrace$.\qed

\vspace{5mm}
\noindent \textbf{Proof of Lemma \ref{avgcost}.}
At each iteration, the probability that a pattern $P_i$ leads to the verification step is $\frac{1}{b!}$ and the cost for the verification for $P_i$ is $O(|P_i|)$ by Lemma~\ref{lemma:nn matching}. Since there are $k$ patterns, the expected cost of the verification step at each iteration is $\sum_{i=1}^{k}\frac{O(|P_i|)}{b!}=\frac{O(M)}{b!}=O(1)$.\qed

\vspace{5mm}
\noindent \textbf{Proof of Lemma \ref{lemma6}.}
At each iteration, the probability that a pattern $P_i$ leads to the verification is at most $s^{(m-1)/2}$. Thus, the expected cost of the verification step is at most $\sum_{i=1}^{k}s^{(m-1)/2}\cdot O(|P_i|)= O((\frac{5}{8})^{m/2}\cdot M)$, which is $O(1)$ when $M$ is polynomial with respect to $m$.\qed

\begin{figure}
  \begin{center}
  \includegraphics[scale=0.45]{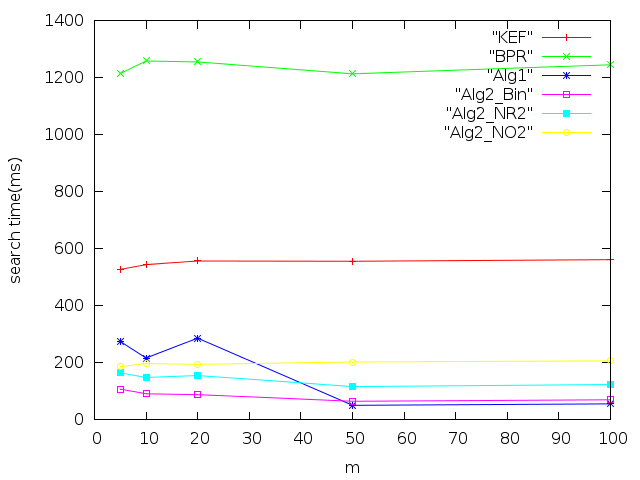}
  \caption{Average search times when $k=10$.}
  \label{NumberOfPattern10}
  \end{center}
\end{figure}

\begin{figure}
  \begin{center}
  \includegraphics[scale=0.45]{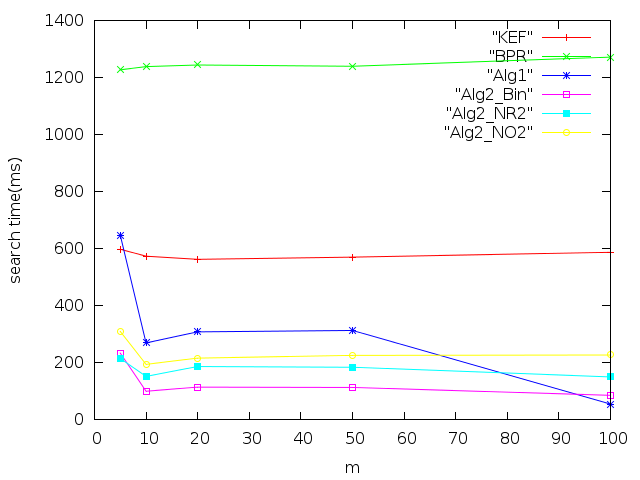}
  \caption{Average search times when $k=50$.}
  \label{NumberOfPattern50}
  \end{center}
\end{figure}

\begin{figure}
  \begin{center}
  \includegraphics[scale=0.45]{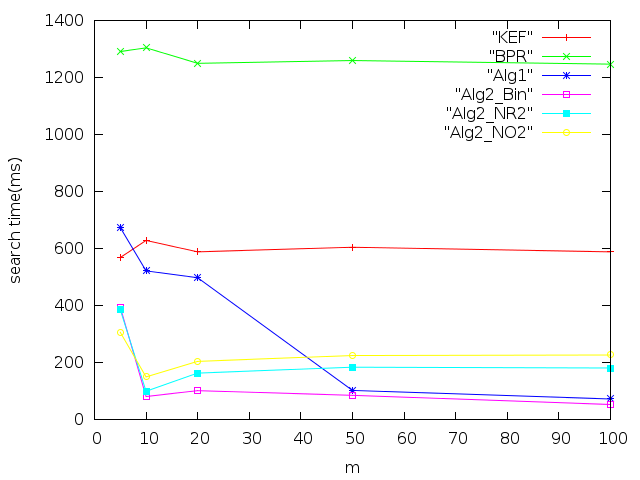}
  \caption{Average search times when $k=100$.}
  \label{NumberOfPattern100}
  \end{center}
\end{figure}

\end{document}